\documentclass[letterpaper,11pt]{article}

\usepackage{scrextend}
\usepackage{amsmath,amssymb,epsfig,bbm}
\usepackage[active]{srcltx}
\usepackage[all]{xypic}
\usepackage{MnSymbol}
\usepackage{slashed}
\usepackage{amsthm}

\usepackage{color}

\usepackage{dsfont}

\definecolor{co}{cmyk}{0,0.7,0.3,0}
\definecolor{darkgreen}{cmyk}{1,0,1,.2}
\definecolor{m}{rgb}{1,0.1,1}

\newcommand{\be}{\begin{equation}}
\newcommand{\ba}{\begin{eqnarray}}
\newcommand{\ea}{\end{eqnarray}}
\newcommand{\nn}{\nonumber}

\def\d{\delta}
\def\e{\epsilon}

\def\m{\mu}
\def\n{\nu}

\def\r{\rho}

\def\OO{\Omega}

\def\ca{{\cal A}}
\def\cb{{\cal B}}

\def\cf{{\cal F}}

\makeatletter
\newcommand{\eqnum}{\refstepcounter{equation}\textup{\tagform@{\theequation}}}
\makeatother

\newcommand{\C}{{\Bbb C}}

\newtheorem{thm}{Theorem}[subsection]

\newtheorem{definition}[thm]{Definition}

\newtheorem*{definition*}{Definition}

\fontfamily{yfrak}

\begin{document}

\vskip 25mm

\begin{center}

{\large\bfseries  
Quantum Holonomy Theory and\\ Hilbert Space Representations
}


\vskip 6ex

Johannes \textsc{Aastrup}\footnote{email: \texttt{aastrup@math.uni-hannover.de}} \&
Jesper M\o ller \textsc{Grimstrup}\footnote{email: \texttt{jesper.grimstrup@gmail.com}}\\ 
\vskip 3ex  


\end{center}

\vskip 3ex

\begin{abstract}


\end{abstract}

We present a new formulation of quantum holonomy theory, which is a candidate for a non-perturbative and background independent theory of quantum gravity coupled to matter and gauge degrees of freedom. The new formulation is based on a Hilbert space representation of the $\mathbf{QHD}(M)$ algebra, which is generated by holonomy-diffeomorphisms on a 3-dimensional manifold and by canonical translation operators on the underlying configuration space over which the holonomy-diffeomorphisms form a non-commutative $C^*$-algebra. A proof that the state that generates the representation exist is left for later publications.

\newpage
\tableofcontents

\section{Introduction}
\setcounter{footnote}{0}

At the heart of modern theoretical physics lies the ancient question: "{\it does a fundamental principle, that explains all of reality and that cannot itself be reduced to other, deeper principles, exist?}  This is the search for a {\it final} theory. 

With {\bf quantum holonomy theory} we propose a candidate for such a fundamental principle and a candidate for a final theory. Quantum holonomy theory is based on an elementary $\ast$-algebra called the {\it quantum holonomy-diffeomorphism algebra} -- denoted by $\mathbf{Q H D} (M)  $ -- which is generated first by holonomy-diffeomorphisms on a 3-dimensional manifold $M$ and secondly by canonical translation operators on an underlying configuration space of connections -- the space over which the holonomy-diffeomorphisms form a non-commutative $C^*$-algebra of functions \cite{Aastrup:2014ppa}.

Thus, the $\mathbf{QH D} (M)  $ algebra simply encodes how {\it stuff} is moved around in a 3-dimensional manifold. This algebra is so elementary -- and conceptually almost empty -- that it seems like the perfect foundation for a {\it final} theory. The question "{\it what are diffeomorphisms made of?} makes little sense.

The theory, that the $\mathbf{QH D} (M)  $ algebra gives rise to, has several key characteristics of a non-perturbative and background independent theory of quantum gravity coupled to matter and gauge degrees of freedom. 
First of all, an infinitesimal version of the central algebraic relation in the $\mathbf{QH D} (M)  $ algebra reproduces the canonical commutation relations of canonical quantum gravity formulated in terms of Ashtekar variables\footnote{In the present setup we use the gauge group $SU(2)$ which in the canonical setup corresponds to a Euclidean signature \cite{AL1}.} \cite{Ashtekar:1986yd,Ashtekar:1987gu}. This means that a Hilbert space representation of the $\mathbf{QH D} (M)  $ algebra will automatically involve information about the kinematics of quantum gravity \cite{Aastrup:2015gba}. 
Secondly, the algebra of holonomy-diffeomorphisms produces in a semi-classical limit -- based on a semi-classical state -- an {\it almost commutative algebra} \cite{Aastrup:2012vq}, i.e. the key ingredient in the non-commutative formulation of the standard model of particle physics coupled to general relativity \cite{Connes:1996gi,Chamseddine:2007hz} (see \cite{Boyle:2016cjt} for an interesting recent development). This hints at  a possible connection to the standard model itself.\\

The first question one is faced with when building a theory over the $\mathbf{Q H D} (M)  $ algebra is what Hilbert space representations it has. This question was central to the papers \cite{Aastrup:2014ppa,Aastrup:2015gba,Aastrup:2012vq,Aastrup:2016ytt}. There we first used lattice approximations and later a lattice independent formulation to analyse the question. What we discovered was, however, a simple "no state" argument that seemed to rule out states on $\mathbf{Q H D} (M)  $ -- a finding that made us consider alternative formulations that involved a Dirac type operator and a certain flow-dependent formulation of the $\mathbf{Q H D} (M)  $ algebra \cite{Aastrup:2015gba,Aastrup:2016ytt}. 

In this paper we argue, however, that there exist a way around the "no state" argument.  The "no state" argument was based on a functoriality condition with respect to the manifold of the state acting on the translation operators. We argue that it is possible to have states on $\mathbf{Q H D} (M)  $, which are physically realistic and which do not satisfy this functoriality condition. These states will then provide us with a kinematical Hilbert space via the GNS construction.



The outline of this paper is as follows.  In section 2 we introduce the algebra generated by holonomy-diffeomorhisms, denoted $\mathbf{H D} (M)  $, and its extension $\mathbf{QH D} (M)  $ that also include the translation operators over the underlying configuration space of connections. In section 3 we then construct a candidate for a state. Next we link the construction to canonical quantum gravity in section 4 by showing that the canonical commutation relations of canonical quantum gravity formulated in terms of Ashtekar variables emerges from the key algebraic relation of the $\mathbf{QH D} (M)  $ algebra.  Finally,  we end in section 5 with a discussion.

\section{The Quantum holonomy-diffeomorphism algebra}
\label{firsttask}

We start with the holonomy-diffeomorphism algebra $\mathbf{H D} (M)  $, which was first introduced in \cite{Aastrup:2012vq}, and the quantum holonomy-diffeomorphism algebra $\mathbf{QH D} (M) $  as well as its infinitesimal version $\mathbf{dQH D} (M)  $, which  were introduced in \cite{Aastrup:2014ppa} and \cite{Aastrup:2015gba}.

\subsection{The holonomy-diffeomorphism algebra}
\label{beent}

Let $M$ be a compact and connected $3$-dimensional manifold. Consider the vector bundle $S=M\times \C^2$ over $M$ as well as the space of $SU(2)$ connections acting on the bundle. Given a metric $g$ on $M$ we get the Hilbert space $L^2(M,S,dg)$, where we equip $S$ with the standard inner product. Given a diffeomorphism $\phi:M\to M$ we get a unitary operator $\phi^*$ on  $L^2(M,S,dg)$ via
$$( \phi^* (\xi ))(\phi (x) )= (\Delta \phi )(M)  \xi (x) , $$
where  $\Delta \phi (x)$ is the volume of the volume element in $\phi (x)$ induced by a unit volume element in $ x$ under $\phi $.      

Let $X$ be a vectorfield on $M$, which can be exponentiated, and let $\nabla$ be a $SU(2)$-connection acting on $S$.  Denote by $t\to \exp_t(X)$ the corresponding flow. Given $x\in M$ let $\gamma$ be the curve  
$$\gamma (t)=\exp_{t} (X) (x) $$
running from $x$ to $\exp_1 (X)(x)$. We define the operator 
$$e^X_\nabla :L^2 (M , S, dg) \to L^2 (M ,  S , dg)$$
in the following way:
we consider an element $\xi \in L^2 (M ,  S, dg)$ as a $\C^2$-valued function, and define 
\begin{equation}
  (e^X_\nabla \xi )(\exp_1(X) (x))=  ((\Delta \exp_1) (x))  \hbox{Hol}(\gamma, \nabla) \xi (x)   ,
  \label{chopin1}
 \end{equation}
where $\hbox{Hol}(\gamma, \nabla)$ denotes the holonomy of $\nabla$ along $\gamma$. 
Let $\ca$ be the space of $SU(2)$-connections. We have an operator valued function on $\ca$ defined via 
\begin{equation}
\ca \ni \nabla \to e^X_\nabla  . 
\nn
\end{equation}
We denote this function $e^X$. For a function $f\in C^\infty_c (M)$ we get another operator valued function $fe^X$ on $\ca$. We call this operator a holonomy-diffeomorphisms. 
Denote by $\cf (\ca , \cb (L^2(M, S,dg) ))$ the bounded operator valued functions over $\ca$. This forms a $C^*$-algebra with the norm
$$\| \Psi \| =  \sup_{\nabla \in \ca} \{\|  \Psi (\nabla )\| \}, \quad \Psi \in  \cf (\ca , \cb (L^2(M, S,dg )) ). $$

\begin{definition}
Let 
$$C =   \hbox{span} \{ fe^X |f\in C^\infty_c(M), \ X \hbox{ exponentiable vectorfield }\}  . $$
The holonomy-diffeomorphism algebra $\mathbf{H D} (M,S,\ca)   $ is defined to be the $C^*$-subalgebra of  $\cf (\ca , \cb (L^2(M,S,dg )) )$ generated by $C$.
We will often denote $\mathbf{H D} (M,S,\ca)   $ by  $\mathbf{H D}  (M)$ when it is clear which $S$ and $\ca$ is meant.
\end{definition}

It was shown in \cite{AGnew} that  $\mathbf{H D} (M,S,\ca)   $ is independent of the metric $g$. \\

\subsection{The quantum holonomy-diffeomorphism algebra}

Let $\mathfrak{su}(2)$ be the Lie-algebra of $SU(2)$.   
A section $\omega \in \Omega^1(M,\mathfrak{su}(2))$ induces a transformation of $\ca$, and therefore an operator $U_\omega $ on $\mathcal{F}(\ca,  \cb(L^2 (M ,  S,g)))$ via   
$$U_\omega (\xi )(\nabla) = \xi (\nabla - \omega) ,$$ 
which satisfy the relation 
\begin{equation} \label{konj}
(U_{\omega} fe^X U_{ \omega}^{-1}) (\nabla) = fe^X (\nabla + \omega )  .
\end{equation}
%
%
Infinitesimal translations on $\ca$ are given by 
\begin{equation}
E_\omega  =\frac{d}{dt}U_{  t  \omega}\Big|_{t=0} \;,
\label{soevnloes}
\end{equation}
where we note that 
$$
E_{\omega_1+\omega_2}=E_{\omega_1}+E_{\omega_2\;,}
$$
which follows since the map $\Omega^1 (M,\mathfrak{su}(2))\ni \omega \to U_{ \omega}$ is a group homomorphism, i.e. $U_{(\omega_1+\omega_2 )}=U_{\omega_1}U_{ \omega_2}$. 

We define the $\mathbf{QHD}(M)$ as the algebra generated by elements in $\mathbf{HD}(M)$ and by translations $U_{\omega}$. We define the infinitesimal quantum holonomy-diffeomorphism algebra $\mathbf{dQHD}(M)$ as the algebra generated by elements in $\mathbf{HD}(M)$ and by infinitesimal translations $E_{\omega}$.\\

Elements of $\mathbf{QHD}(M)$ can be written in a canonical form. To see this we first define
$$
e^X_\omega := U^{-1}_\omega e^X U_\omega
$$
which permits us to write any combination of $U_\omega$'s and $fe^X$'s in the canonical form
$$
 U_{\omega_1} f_2 e^X_{\omega_2}\ldots f_n e^X_{\omega_n}
$$
for some $\omega_i\in \Omega^1(M,\mathfrak{su}(2))$.  
Note that we have the relations 
$$
e^X_{\omega_1} U_{\omega_2} = U_{\omega_2} e^X_{\omega_1 + \omega_2}
$$
and 
$$
\left( U_{\omega_1} e^X_{\omega_2} \right)^* = U_{-\omega_1} e^{-X}_{-\omega_1 + \omega_2}\;.
$$

\section{States on $\mathbf{QHD}(M)$}

We will now consider states on $\mathbf{QHD}(M)$. 
We think of a state as a map
\begin{equation}
\rho: \mathbf{QHD}(M)  \rightarrow M_2(\mathbb{C})\otimes \cf(M\times M)
\label{morgensang}
\end{equation}
where we obtain the actual state by composing $\rho$ with the map
\begin{eqnarray}
&&\Phi_{\psi}: M_2(\mathbb{C})\otimes \cf(M\times M) \rightarrow \mathbb{C} , 
\nn\\
&&\hspace{0.8cm} K(x,y)   \rightarrow   \int_{M\times M}     \bar{\psi}(x) K(x,y)  \psi(y)dxdy,
\label{thy}
\end{eqnarray}
where $\psi$ is a $\mathbb{C}^2$-valued half-density on $M$. Alternatively a 'vacuum' state can be obtained via
\begin{equation}
K(x,y)   \rightarrow   \int_{M\times M}   Tr_{M_2}  K(x,y)\d^{3}(x-y) dxdy,
\label{thedoors}
\end{equation}
where it is implicit understood that we integrate over appropriate half-densities.

Let ${\bf A}$ be a map from $TM$ to $M_2(\C)$, whose properties will be specified shortly. 
We specify the map $\rho_{\bf A}^\kappa$, which is to be a state on $\mathbf{QHD}(M)$, first with 
\begin{eqnarray}
\rho_{\bf A}^\kappa( f e^X_\omega) (x,y) = f(x) \hbox{Hol}(\gamma,{\bf A}+\omega) \d^{3}(y-\exp(X)(x)) 
\label{hvisker1}
\end{eqnarray}
where $\gamma$ is the curve in $M$ generated by $X$ and which connects $x$ and $y$. The general expression reads
\begin{eqnarray}
\rho_{\bf A}^\kappa(    e^{X_1}_{\omega_1}\ldots  e^{X_n}_{\omega_n}   ) (x,y) 
&=&  \hbox{Hol}(\gamma_2,{\bf A}+\omega_2)\ldots \hbox{Hol}(\gamma_n,{\bf A}+\omega_n) 
\nn\\&& 
\cdot\d^{3}(y-\exp(X)(x)) 
\end{eqnarray}
where $\gamma_i$ is the path generated by the vector field $X_i$. We write $X$ as a shorthand for the combination of all the involved vector fields.
We then write the state as
\begin{eqnarray}
\rho^\kappa_{(\psi,{\bf A})}=\Phi_{\psi}\circ \rho_{\bf A}^\kappa\;.
\label{tristese}
\end{eqnarray}
Next, we need to specify the state on operators, which have the general structure
$
 U_{\omega_1} e^{X_2}_{\omega_2}\ldots  e^{X_n}_{\omega_n}
$. 
There appears to be two ways to construct such a state. The first is to write
\begin{eqnarray}
\hspace{-1cm}1)\quad\rho_{(\psi,{\bf A})}^\kappa(   U_{\omega_1} e^{X_2}_{\omega_2}\ldots  e^{X_n}_{\omega_n}   ) 
= \OO_{\bf A}^\kappa(  \omega_{1})   \rho_{(\psi,{\bf A},{\bf E})}^\kappa  (   e^{X_2}_{\omega_2}\ldots  e^{X_n}_{\omega_n}   ) 
\label{NOET}
\end{eqnarray}
where $ \OO_{\bf A}^\kappa(  \omega) $ is a function that satisfies
$$
0\leq\left\vert \OO_{\bf A}^\kappa(\omega) \right\vert \leq1\;,\quad \OO_{\bf A}^\kappa(0)=1\;.
$$
The second way to construct a state is to let $\OO_{\bf A}^\kappa$ be also a function on $M$. In the following we shall use the same symbol $ \OO_{\bf A}^\kappa $ for both cases and trust that no confusion will arise. For the second option we write
\begin{eqnarray}
\hspace{-1,1cm}2)\quad\rho_{\bf A}^\kappa(   U_{\omega_1} e^{X_2}_{\omega_2}\ldots  e^{X_n}_{\omega_n}   ) (x,y) 
&&\nn\\
&&\hspace{-2cm}= \OO_{\bf A}^\kappa(  \omega_{1}) (x)  \rho_{\bf A}^\kappa  (   e^{X_2}_{\omega_2}\ldots  e^{X_n}_{\omega_n}   ) (x,y)
\label{NOTO}
\end{eqnarray}
and then construct the state via (\ref{tristese}). If we employ the latter method it is the integral of $\vert\OO_{\bf A}^\kappa(\omega)\vert$ that must lie between zero and one and equal one when evaluated on $\omega=0$.

Regardless of which way the candidate for a state is constructed the function $\OO_{\bf A}^\kappa $ is the key element to understand it. But before we turn our attention hereto let us first consider the element
${\bf A}$.
We write ${\bf A}$ as 
$$
{\bf A} = \mathds{A} + \mathds{A}_q
$$
where $\mathds{A}$ is a one-form, which takes values in $\mathfrak{su}(2)$. 
$ 
\mathds{A}_q
$ 
is a map $TM\rightarrow M_2(\mathbb{C})$ that satisfies the following homogeneity condition
$$
\mathds{A}_q(\lambda X) = \vert \lambda \vert \mathds{A}_q( X)\;,\quad \lambda\in  \mathbb{C}
$$
and where $\mathds{A}_q( X) $ is a negative definite element in $M_2(\mathbb{C})$.
Here $\kappa$ enters as a quantization parameter, which separates the classical contribution from its 'quantum' counterparts.

The idea is  to interpret $\mathds{A}$ as an Ashtekar connection. The conjugate Ashtekar variables, the densitised triad field, must then emerge from $\OO_{\bf A}^\kappa$ in the semi-classical limit.

Let us now return to the function $\OO_{\bf A}^\kappa$ introduced in (\ref{NOET}) and (\ref{NOTO}). If we first consider the second option, then the function $\OO_{\bf A}^\kappa$ cannot be a local function on $M$ of $\omega$, since this would bring the state in conflict with the $\mathbf{HD}(M)$ algebra. The problem being the following: an $\omega$ which is zero on  $M$ apart from a  region with small volume, would have $\rho^k_{(\psi,{\bf A})}(U_\omega)$ almost equal to one, i.e. if one considers $\mathbb{A}$ or $\mathbb{A}+\omega$ then they would basically define the same state. However since the holonomy of the flows only involve a one-dimensional integral, these integrals could be "big" running through that region,  i.e. the holonomy on  $\mathbb{A}$ and $\mathbb{A}+\omega$ would be very different contradicting that $\rho^k_{(\psi,{\bf A})}(U_\omega)$ is almost equal to one.  
Also when  $\OO_{\bf A}^\kappa$ were local, it satiesfy properties  similar to the one used in  the argument made in section 6 in \cite{Aastrup:2016ytt}, which lead to the conclusion that under these assumptions there cannot be a non-trivial state on the $\mathbf{QHD}(M)$ algebra. One of the assumptions made in this argument was that $\rho_{(\psi,{\bf A})}^\kappa(   U_{\omega} )$ behaves functorial with respect to restriction of $\omega$ on $M$. 
If, however, $\OO_{\bf A}^\kappa(\omega)$ is not a local function, then this argument does not apply. This observation also applies to the first option (\ref{NOET}), where it implies that $\OO_{\bf A}^\kappa$ cannot be an integral over $M$ of a local function. 

This suggests that $\OO_{\bf A}^\kappa(\omega)$ involves derivations of $\omega$ and that it will effectively introduce a kind of spatial cut-off that involves a scale. This will then ensure that $\OO_{\bf A}^\kappa$ does not behave functorially with respect to $M$ on short scales.

Note that the physically interesting state seem to arise only from the second of the above options, where $\OO_{\bf A}^\kappa$ is a function on $M$ that interacts with the holonomy-diffeomorphisms. The reason for this is that the conjugate variables -- as we shall show in the next section -- i.e. connections and inverse triad fields, come from the holonomy-diffeomorhisms and the translation operators respectively. If these two variables do not interact within the same integral it seems impossible to form operators, which correspond to physical quantities such as the Dirac and gravitational Hamiltonians.

 It is beyond the scope of this paper to analyse what condition $ \OO_{\bf A}^\kappa$ must satisfy in order to have a state. We suspect that the definition of infinitesimal operators and the computation of the constraint algebra and the requirement to have off-shell closure\footnote{see \cite{Aastrup:2015gba} for computations in the lattice formulation.} will have an impact on what $ \OO_{\bf A}^\kappa $ may look like. We shall address this key issue in a later publication.

\section{Connection to canonical quantum gravity}
\label{session}

Let us end with a section on the connection between the $\mathbf{QHD}(M)$ algebra and canonical quantum gravity. The following is a rehash of material published in \cite{Aastrup:2014ppa}.

If we combine equation (\ref{konj}) with (\ref{soevnloes}) we obtain
\begin{equation} \label{flowkan}
[ E_\omega , e^X_\nabla ]= \left.\frac{d}{dt}e^X_{\nabla +t\omega}\right\vert_{t=0}  . 
\end{equation}
To analyse the righthand side of (\ref{flowkan}) we introduce local coordinates $(x_1 ,x_2,x_3)$ and write $\omega =\omega^i_\m\sigma_i dx^\m$. 
For a given point $p\in M$ choose the points $$p_0=p,\quad p_1=e^{\frac{1}{n}X}(p),\ldots  ,\quad  p_n= e^{\frac{n}{n}X}(p)$$ 
on the path
$$t\to e^{tX}(p)  ,t\in [0,1].$$
We write the vectorfield $X=X^\n\partial_\n$. 
We have 
\begin{eqnarray} 
\lefteqn{e^X_{\nabla+t\omega}}\nn\\
& =&\lim_{n\to \infty } (1+\frac{1}{n}(A(X(p_0))+t \omega^i_\m \sigma_i X^\m(p_0) ) (1+\frac{1}{n}(A(X(p_1))+t \omega^i_\m \sigma_iX^\m(p_1))\nn\\
&& \cdots (1+\frac{1}{n}(A(X(p_n))+t \omega^i_\m \sigma_i X^\m(p_n)) ,
\label{vlad}
\end{eqnarray}
where  $\nabla=d+A$, and therefore 
\begin{eqnarray}
 \lefteqn{\frac{d}{dt}e^X_{\nabla +t\omega}\Big\vert_{t=0}}
 \nn
 \\
 &=& \lim_{n\to \infty }  \Big( \frac{1}{n} \omega^i_\m\sigma_iX^\m(p_0)  (1+\frac{1}{n}A(X(p_1)))\cdots (1+\frac{1}{n}A(X(p_n))) \nn\\
 &&+ (1+\frac{1}{n}A(X(p_0))) \frac{1}{n}  \omega^i_\m \sigma_iX^\m(p_1)  (1+\frac{1}{n}A(X(p_2)))\cdots (1+\frac{1}{n}A(X(p_n))) \nn\\
 && + \quad\quad\quad\quad\quad\quad\quad\quad\quad\quad\quad \vdots \nn\\
 &&+ (1+\frac{1}{n}A(X(p_0)))   (1+\frac{1}{n}A(X(p_2)))\cdots 
 \nn\\
 &&\hspace{4cm}\cdots
 (1+\frac{1}{n}A(X(p_{n-1}))) \frac{1}{n}  \omega^i_\m \sigma_i X^\m(p_n) \Big).
\label{COOM}
\end{eqnarray}
If we restrict this to a path $\gamma(t)$, $t\in [a,b]$, generated by the vector field $X$ equation (\ref{COOM}) gives a line integral of the form
\begin{equation}
\int dt Hol ({\gamma_{<t} },\nabla)  \omega (\dot{\gamma}(t)) Hol( {\gamma_{>t} },\nabla)  
\label{omega}
\end{equation}
where $\gamma_{<t}$ is the path $[a,t] \ni \tau\to \gamma(\tau)$ and where $\gamma_{>t}$ is the path $[t,b] \ni \tau\to \gamma(\tau)$. 

Equation (\ref{COOM}) and (\ref{omega}) should be compared to the classical setup of Ashtekar variables and holonomies of Ashtekar connections \cite{AL1}. There we have canonically conjugate variables $(\mathds{E}^\m_i , \mathds{A}_\n^j)$ where indices $\{i,j,k,...\}$ are $\mathfrak{su}(2)$ indices and $\{\m,\n,...\}$ are indices labelling a coordinate system on $M$. $\mathds{E}$ is a densitized inverse triad field $\mathds{E}^\m_i= e e^\m_i$ where $e^\m_i$ is the inverse triad field and $e$ its determinant. $\mathds{A}$ is the Ashtekar connection\footnote{Note again that we here work with $SU(2)$ connections which in a canonical framework correspond to either a Euclidian signature or a Hamiltonian with a comparably more complicated structure, see for instance \cite{AL1}. }. If one considers instead of $\mathds{E}$ its flux over a two-surface $S$
\begin{equation}
F^S_i =\int_S \e_{\m\n\r}  \mathds{E}_i^\m dx^\n dx^\r
\label{flux}
\end{equation}
then the Poisson bracket between the holonomy of $\mathds{A}$ along a curve $\gamma$ and $F^S_i$ reads \cite{AL1}
$$
\left\{  F^S_i, Hol(\gamma,\mathds{A})   \right\}_{PB} = \iota(S,\gamma)  Hol(\gamma_1,\mathds{A}) \sigma_i Hol(\gamma_2,\mathds{A})  
$$
where $\gamma = \gamma_1 \cdot \gamma_2$ and where the Pauli matrix is inserted at the point of intersection between $S$ and $\gamma$. $\iota(S,\gamma)=\pm 1$ or $0$ encodes information on the intersection of $S$ and $\gamma$.   

We therefore see that before taking the limit $\lim_{n\to \infty}$ of (\ref{COOM}) we have simply the commutator of the sum of the flux operators $\sum_k \frac{1}{n}X^\m (p_k) F^{S_k}_i $, where $S_k$ is the plane orthogonal to the $x_\m$-axis intersecting $p_k$, and the holonomy operator of the path $$t\to e^{tX}(p)  ,t\in [0,1].$$
It follows that $E_{\sigma_i dx^\m}$ is  a series of flux-operators $F^S_i$ sitting along the path $$t\to e^{tX}(p)  ,t\in [0,1],$$
where the surfaces $S$ are just the planes othogonal to the $x_\m$ direction.
But since there are infinitely many of them, they have been weighted with the infinitesimal length, i.e. with a $dx^\m$.
We can formally write this as
$$
E_{\sigma_i dx^\m} = \int_M \hat{F}^S_i dx^\m  \;, 
$$
where $ \hat{F}^S_i$ is an operator, which corresponds to a quantization of the flux operator (\ref{flux}).   
This provides us with a solid interpretation of the $\mathbf{QHD}(M)$ algebra in terms of canonical quantum gravity, where the operator $E_\omega$ is a global flux operator.



Note, however, that $E_\omega$ transforms as a one-form. This means that $E_\omega$ will {\it not}, when we incorporate a classical correspondence via a state, correspond to the Ashtekar variable $\mathds{E}^\m_i$ but rather to the inverse triad field $e^\m_i$. The reason for this is that $\mathds{E}$ is a densitised field and thus does not transform as a one-form. This corresponds to the fact that the determinant $e$ emerges in our framework via the state, which is based on half-densities on $M$. This, in turn, implies that we are dealing with a non-canonical framework.

Finally, we can also make the holonomies infinitesimal in order to see the canonical commutation relations between the Ashtekar variables directly. This was done in \cite{Aastrup:2015gba} and shall not be done here.

\section{Discussion}

We have presented an alternative formulation of {\it quantum holonomy theory} based on a Hilbert space representation of the $\mathbf{QHD}(M)$ algebra. In previous publications we believed that such representations were ruled out by a certain functoriality condition, but in these pages we argue that there exist a way to construct physically interesting states, that do not satisfy this condition.
The result will be a more natural formulation of the theory, where it is the $\mathbf{QHD}(M)$ algebra itself -- and not an infinitesimal and flow-dependent variant thereof \cite{Aastrup:2016ytt} -- that provides the quantum variables.

It remains, however, to analyse whether such a state in fact exist. This comes down to analysing what conditions $ \OO_{\bf A}^\kappa$ must satisfy and whether such conditions can be met. We expect that the required non-locality of $ \OO_{\bf A}^\kappa$ will introduce a scale and possible a whole sequence of free parameters. We also expect that the formulation of infinitesimal operators with a correct semi-classical limit will further restrict the form of $ \OO_{\bf A}^\kappa$. 

Note the difference between the $\mathbf{HD}(M)$ and $\mathbf{QHD}(M)$ algebra with respect to gauge transformations. The spectrum of the $\mathbf{HD}(M)$ algebra is given by certain connections modulus unitary transformations, i.e. gauge transformations. The $\mathbf{QHD}(M)$ algebra, on the other hand, involves gauge transformations. We suspect that this difference will play a role when analysing whether a state exist on $\mathbf{QHD}(M)$.

This key issue aside, an interesting aspect of quantum holonomy theory is the way it deals with the square root of the determinant of the metric. The conjugate Ashtekar variable is an inverse {\it densitized} triad field, which means -- when incorporated in a quantum theory -- that one has to deal with various powers of an operator that corresponds to the square root of the determinant of the metric, something that causes serious difficulties. In the present approach the square root of the determinant of the metric arises not from the quantum operators but from the state via integrals over half-densities. This implies first of all that quantum holonomy theory departs from a canonical quantization scheme. It is something very similar, but nevertheless different. Secondly it means that the construction of operators, that correspond for example to the Hamilton and diffeomorphism constraints, should be much simpler.

Finally, a key aspect of the theory, which we would like to emphasize, is that the state, over which the kinematical Hilbert space is build, is semi-classical. This means that this approach to a theory of quantum gravity {\it automatically} includes a semi-classical limit. This aspect is in great contrast to other approaches to non-perturbative approaches to quantum gravity \cite{AL1}, where the semi-classical limit is a serious challenge.


\end{document}